\documentclass[a4paper]{article}
\usepackage{multicol}
\usepackage{amsmath}
\usepackage{mathrsfs}
\usepackage[utf8]{inputenc}
\usepackage{hyperref}
\usepackage{cite}
\usepackage{graphicx}
\usepackage[left=20mm, top=20mm, right=20mm, bottom=20mm, nohead, nofoot]{geometry}
\usepackage{caption}

\hypersetup{pdfstartview=FitH}

\newcommand{\s}[1]{\hat \sigma_{#1}}
\newcommand{\vF}{v_\text{\tiny F}}
\newcommand{\qe}{\varepsilon_\text{eff}}
\newcommand{\bse}{\tilde\varepsilon}
\newcommand{\p}[1]{\hat \pi_{#1}}
\newcommand{\B}[3]{J_{#1}^{#2}\left(#3\right)}
\newcommand{\com}{\mspace{5mu}\text{,}}
\newcommand{\pnt}{\mspace{5mu}\text{.}}
\newcommand{\En}{\hat{\mathscr H}}
\begin{document}
\onecolumn
\begin{center}
	\large{\textbf{Magnetic Minibands and Electron-Electron Bound States in ac-Driven Graphene with Space-Modulated Gap}}
\end{center}
\vspace{1ex}
\begin{center}
	S.V. Kryuchkov $^\text {a, b}$, E.I. Kukhar $^\text {a, }$\footnote{ eikuhar@yandex.ru}
\end{center}
\vspace{1ex}
\begin{center}
	$^\text a$\textit{Volgograd State Socio-Pedagogical University, Physical Laboratory of Low-Dimensional Systems\footnote{ \url{http://edu.vspu.ru/physlablds}}, V.I. Lenin Avenue, 27, Volgograd 400066, Russia}
\vspace{2ex}
	
	$^\text b$\textit{Volgograd State Technical University, V.I. Lenin Avenue, 28, Volgograd 400005, Russia}
\end{center}

\begin{abstract}
The emergence of the magnetic minibads in the quasienergy spectrum of graphene superlattice subjected to the quantizing magnetic field and electromagnetic radiation was investigated.
The graphene superlattice was assumed to be formed by space-periodical modulation of the gap in the vicinity of the Dirac point of the graphene band structure.
The explicit form of the Floquet spectrum of electron was derived in the case of weak spatial modulation. Minibads widths were shown to change with ac electric field amplitude.
The possibility of electron-electron bound states was shown. The binding energies of these states were shown to be the function of the amplitude of ac electric field.
\\
\\
\textbf{Keywords}: \textit{graphene superlattice; magnetic minibands; Floquet spectrum; quasienergy; bound states}
\\
\\
\\
\end{abstract}

\begin{multicols}{2}

\textbf{I. Introduction}\\

Modern technologies of fabrication of graphene-like materials lead to the new area for investigations of 2D Dirac-like fermions \cite{1,2,3,4}.
Presently graphene is offered to be the basis of the modern nano- and microelectronics \cite{5,6,7,8,9,10,11}.
In \cite{12,13} the use of graphene for generation of electromagnetic (EM) radiation in THz range had been offered.
From an application-technological point of view the tunability of electronic and optical properties of graphene materials by the external EM fields has particular importance.
This gives the way of manipulation of spectrum parameters without interfering in the internal structure of graphene material \cite{14,15,16,17,18,19,20,21}.
The latter is important for the engineering of devices with tunable characteristics \cite{22}.

Electronic properties of graphene-like materials can be tuned by time-dependent driving (or ac-driving) into specific states called as the Floquet topological insulators \cite{23,24}.
Such states are possible owing to the dynamical gap which can be induced by EM radiation in the vicinity of the Dirac points of the band structure of the originally gapless graphene layer \cite{25}.
In \cite{26} polarization tunable gaps in the band structure of topological insulator had been observed by using of mid-infrared EM radiation. Floquet topological insulators had been also realized in laboratory in \cite{27}.
Theory describing magneto-electronic properties of ac-driven graphene developed in \cite{28,29,30}.
Results of investigations of band structure of ac-driven graphene allow the manipulation of Dirac points \cite{31}, Landau levels \cite{28,29,30} and states around defects of the graphene lattice \cite{32} by changing of ac-field amplitude.
In \cite{33} the ac electric field had been shown to change the properties of the electron collective excitations which form the plasmon complexes in doped graphene.
So the plasmon frequency was strongly influenced by the external field and the stability of these excitations can be controlled by changing of the ac-field amplitude.

Emergence of bound states (BSs) due to the interactions of electrons with elementary excitations in solid has fundamental and practical significance.
It leads to such many-body phenomena as superconductivity and charge-density waves. Study of interactions of Dirac electrons with elementary excitations and tunable impurities had led to the prediction and observation of new effects in the filed of graphene-like materials \cite{18,19,34,35,36,37,38}.
In this context we can point such phenomena as renormalization of Dirac spectrum due to interactions of electrons with lattice vibrations \cite{39,40}, effect of the electron-phonon coupling on the magnetooptical conductivity \cite{41}, a new structure of magnetophonon resonances determined by the electron-phonon hybrid states formed in the spectrum between the discrete energy levels (Landau levels) appearing in graphene in the strong magnetic field \cite{42,43}, effect of the electron-electron interaction on the graphene resistance and magnetoresistance \cite{44,45}, etc.
In \cite{38} the emergence of macro-molecules due to the screening of Coulomb repulsion between impurity charges in graphene had been investigated.
This screening was provided by the strong magnetic field which localized electrons between impurities so theirs binding could be tuned on or tuned off by magnetic field intensity changing.
Such mechanism of charge bound states had been offered in \cite{38} to manipulate the macro-molecules size.

Investigations of electron-electron interaction are particularly important in the theory of superconductivity of graphene \cite{46,47,48}.
In \cite{46,47} electron correlations had been offered to be a source of the graphene superconductivity.
In [48] an essentially anisotropic electron-electron interaction near the Van Hove singularity had been shown to be responsible for the attraction between electrons in graphene. The possibility of electron-electron coupling in graphene due to the phonon exchanging had been studied in \cite{49}.

Another mechanism of the binding of particles which are characterized by Coulomb repulsion is its transfer (in the quasimomentum space) to the region of spectrum where effective mass is negative \cite{50}.
To this end the use of structure with superlattice (SL) which is characterized by narrow minibands is suitable one \cite{50,51}.
Below we consider the effect of the EM radiation on the possibility of the emergence of electron-electron BS in graphene with SL barriers obtained by the weak space-periodical modulation of the gap in graphene band structure and study the binding-energy dependence on the ac-field amplitude.
\\
\\

\textbf{II. AC-driven graphene with space modulated gap}\\

One of the ways to create the graphene based SL had been offered in \cite{52} where grapheme layer had been proposed to deposit on the periodical substrate. The presence of periodical substrate yields the spatial periodic modulation of the band gap in the graphene.
Below such structure is called as graphene with space-modulated gap (GSMG) and the direction of spatial modulation is called as SL axis.
To investigate the effect of EM radiation on the electron-electron BS the Floquet spectrum of ac-driven GSMG is calculated.

In this section averaging of the Dirac-like Hamiltonian written for the electron in GSMG is produced for the case of arbitrary time-periodical electric signal.
Let the graphene layer is supposed to coincide with plane $xy$.
The profile of the gap space-modulation is given by the function
$\Delta_\text s\left(x\right)=\Delta_\text s\left(x+d\right)$,
where $d$ is the modulation period and $Ox$ is the SL axis.
Besides, GSMG is suggested to be subjected to the time-periodical electric field with intensity oscillating in the plane $xy$.
This can be achieved by placing of the GSMG in the field of EM wave which propagates perpendicular to the GSMG plane $xy$ so that vector of electric field oscillates in $xy$ and is described by the time-dependent vector potential
$\mathbf A^\text{ac}\left(t\right)=\mathbf A^\text{ac}\left(t+T\right)$,
where $T$ is the period of ac-field.

The quantum mechanical state of the electron is described by a spinor $\psi$ which in the vicinity of Dirac point K obeys the Dirac-like equation
$i\hbar\partial_t\psi=\En\psi$.
Here
\begin{gather}\label{2}
\hat {\mathscr{H}}\left(\mathbf{r}{,}t\right)=
\vF \hat {\boldsymbol{\pi}}\cdot \hat {\boldsymbol{\sigma}}
+ \dfrac{e\vF}{c} \mathbf A^\text{ac}\left( t \right)\cdot\hat{\boldsymbol\sigma}
+ \Delta_\text{s} \left( x \right) \s{z} \com
\end{gather}
spinor  $\psi$   has two components corresponding to the different values of pseudospin denoting the graphene sublattice,  $\hat {\boldsymbol{\sigma}}=\left(\s{x}{,}\mspace{15mu} \s{y} \right)$, $ \s{z} $   are the Pauli matrixes, $\vF$   is the Fermi velocity.
Operator $\hat {\boldsymbol{\pi}}$ is related with the operator of momentum $\hat{\mathbf{p}}$ with the sum
$\hat {\boldsymbol{\pi}}=\hat{\mathbf{p}}+\frac{e}{c}\mathbf{A}\left(\mathbf{r}\right) $,
where vector potential $\mathbf{A}\left(\mathbf{r}\right)$ describes the time-independent field. For instance, it may be additional quantizing magnetic field.
Spinor  $\psi$   obeys the Floquet theorem:
\begin{gather}\label{3}
\psi\left(\mathbf{r},t\right)=u\left(\mathbf{r},t\right)e^{-\tfrac{i\qe t}{\hbar}} \com
\end{gather}
where  $ \qe $  is quasienergy,  $ u\left(\mathbf{r},t\right) $   is the spinor with components which are time-periodic functions with period $T$.
The action of the operator (\ref{2}) on the function (\ref{3}) leads to the equation
\begin{gather}\label{4}
\left( \hat {\mathscr{H}}-i\hbar\dfrac{\partial}{\partial t} \right)u=\qe \mspace{5mu} u \pnt
\end{gather}

Polarization of the EM wave is assumed to be linear and we consider two variants of the direction of the ac electric field vector:
(1) the ac-field vector oscillates perpendicularly to the SL axis ($A_x^\text{ac} =0$),
(2) the ac-field is applied along the SL axis ($A_y^\text{ac} =0$).
In both cases to reduce the time-dependent problem described by equation (\ref{4}) to a stationary problem the spinor $u$ is written in the form
$ u\left(\mathbf{r},t\right)=\hat {\mathscr{U}} \chi\left(\mathbf{r},t\right) $.
Here $\hat {\mathscr{U}}$ is the unitary operator which is equal
\begin{gather}\label{14}
\hat {\mathscr{U}}=e^{-i\mspace{2mu}a_j\mspace{2mu}\s{j}}\mspace{5mu} \text{,} \mspace{15mu}
a_j\left( t\right) =\dfrac{e \vF}{c\hbar}\int A_j^\text{ac} \left(t\right) \text{d}t \com
\end{gather}
index $j$ is chosen to be $y$ in the case (1) and it is chosen to be $x$ in the case (2).

Firstly we consider the situation when EM radiation is polarized perpendicularly to the SL axis.
After use of the unitary transformation provided by the operator (\ref{14}), where $j=y$, we arrive at the next equation
\begin{multline}\label{5}
\Bigg(-i\hbar\dfrac{\partial}{\partial t}+\vF \p{y}\s{y}+\\
+\left(\vF \p{x}\s{x}+\Delta_\text{s}\s{z}\right)\hat {\mathscr{U}}^{2}\Bigg)\chi=\qe \mspace{5mu} \chi \pnt
\end{multline}

Further to find the quasienegry appearing in (\ref{5}) we use the averaging method \cite{53} which within our problem is identical to the so-called rotating wave approximation \cite{14}.
To this end we represent the spinor  $\chi$ and time-periodical operator $\hat {\mathscr{U}}^2$  as the sums each contains the constant term and high-frequency term:
$\chi\left(t\right)=\chi_0+\chi_\text{ac}\left(t\right)$,
$\hat {\mathscr{U}}^2\left(t\right)=\hat{\gamma_0}+\hat {\mathscr{U}}_\text{ac}^2$,
where
$$
\chi_0=\begin{pmatrix} \mspace{5mu} b^\uparrow_0 \mspace{5mu} \\ \mspace{5mu} b^\downarrow_0 \mspace{5mu} \end{pmatrix} \text{,}
\mspace{15mu} \chi_\text{ac}\left(t\right)=\sum_{k\neq 0}\begin{pmatrix} \mspace{5mu} b^\uparrow_k \mspace{5mu} \\ \mspace{5mu} b^\downarrow_k \mspace{5mu} \end{pmatrix}
e^{ik\omega t} \mspace{5mu} \text{,} \mspace{15mu}
$$
$$
\hat {\mathscr{U}}_\text{ac}^2=\sum\limits_{k\neq 0} \hat\gamma_k e^{ik\omega t} \mspace{5mu} \text{,} \mspace{15mu}
\omega=2\pi/T \com
$$
\begin{gather} \label{12}
\hat \gamma_k=\dfrac{1}{2\pi}\int\limits_{-\pi}^{+\pi} \hat {\mathscr{U}}^2 \left( \xi \right) e^{-ik\xi}  \text{d}\xi \pnt
\end{gather}
So that the average values are:
$\left<\chi\right>=\chi_0$,
$\left<\chi_\text{ac}\left(t\right)\right>=0$,
$\left\langle \hat {\mathscr{U}}^2\right\rangle =\hat\gamma_0$ and
$\left\langle \hat {\mathscr{U}}_\text{ac}^2\right\rangle =0$.
To neglect the high-frequency part of the equation (\ref{5}) the next conditions are supposed to be performed
\begin{gather}\label{6}
\sum_{k\neq 0}
\left|\Large\mathstrut b^{\uparrow,\mspace{5mu}\downarrow}_k\right|^2
\ll
\left|\Large\mathstrut b_0^{\uparrow,\mspace{5mu}\downarrow} \right|^2  \pnt
\end{gather}
After averaging in the both parts of the equation (\ref{5}) one can find the relation for the constant part
\begin{multline}\label{7}
\left(\hat {\mathscr{H}}_0-\qe\right)\chi_0= \\
=\left(\vF \p{x}\s{x}+\Delta_\text{s}\s{z}\right)
\sum\limits_{k\neq 0}\hat\gamma_k \left\langle\chi_\text{ac} e^{ik\omega t}\right\rangle \com  
\end{multline}
where
\begin{gather}\label{8}
\hat {\mathscr{H}}_0= \vF \p{y}\s{y}+
\left( \vF\p{x}\s{x} +\Delta_\text{s} \s{z}\right) \hat \gamma_0 \com
\end{gather}
Now we leave only the oscillating terms in (\ref{5}).
In which connection the conditions (\ref{6}) allow us to neglect the terms containing $\chi_\text{ac}$ and to arrive at the expression containing only terms with $\chi_0$ and with $\partial_t\chi_\text{ac}$.
After integrating of this expression we obtain
\begin{gather}\label{17}
\chi_\text{ac}\left(t\right) =
-\dfrac{\vF\p{x}\s{x}+\Delta_\text{s}\s{z}}{\hbar\omega}\sum\limits_{k\neq 0}\dfrac{\hat\gamma_k \chi_0}{k} e^{ik\omega t} \pnt
\end{gather}
After substitution of (\ref{17}) into (\ref{7}) we find that the right side of equation (\ref{7}) is zero.
As a result the problem of the electron state in ac-driven GSMG is reduced to the solving of the equation
\begin{gather}\label{10}
\hat {\mathscr{H}}_0\mspace{5mu}\chi_0=\qe\mspace{5mu}\chi_0 \pnt
\end{gather}

The second case suggested here is the situation when EM radiation is polarized along the SL axis ($A_y^\text{ac} =0$).
After calculations which are similar to the above derivation we write instead (\ref{8})
\begin{gather}\label{13}
\hat {\mathscr{H}}_0= \vF \p{x}\s{x}+
\left( \vF\p{y}\s{y} +\Delta_\text{s} \s{z}\right) \hat \gamma_0 \pnt
\end{gather} 
Note that to calculate the matrix $\hat \gamma_0$ appearing in (\ref{13}) we should put into formula (\ref{14}) $j=x$.

Now we consider one particular case.
If the time-independent field described by the vector $\mathbf A\left(\mathbf r\right) $ is absent and gap is not modulated
( $\mathbf A\left(\mathbf r\right)=0 $, $\Delta_\text{s}\equiv\Delta=\text{const}$) then
eigenvalues of the Hamiltonian (\ref{8}) are
\begin{gather} \label{9}
\qe=\pm\sqrt{\vF^2 p_y^2+\left(\alpha_0^2+\beta_0^2\right)\left( \vF^2 p_x^2+\Delta^2\right) } \pnt
\end{gather}
Here sign "$+$" corresponds to the conduction band and sign "$-$" to the valence band, $\alpha_0$ and $\beta_0$ are the real functions of ac-field amplitude and are determined from the formula
$$
\alpha_0+i\beta_0=\dfrac{1}{\pi}\int\limits_{-\pi}^{+\pi}e^{i\mspace{2mu}2a_y\left( \xi \right) }\text{d}\xi \pnt
$$
For the sinusoidal radiation with amplitude of electric field intensity $E_0$ and frequency $\omega$ and for the originally gapless graphene ($\Delta=0$) formula (\ref{9}) gives the result \cite{30}:
$\qe=\pm\sqrt{u_\text{\tiny F}^2 p_x^2+ \vF^2 p_y^2 } \pnt$
Here $u_\text{\tiny F}=\vF\B{0}{}{2a_0}$ is the Fermi velocity renormalized by the ac-field effect, $a_0 = e\vF E_0/\hbar \omega^{2} $ is the dimensionalless amplitude of ac-field, $\B{k}{}{\xi}$ is the Bessel function of integer order.
The approximation provided by the conditions (\ref{6}) is seen from the formula (\ref{17}) to  work correctly if the next inequality is performed
$$
\dfrac{\Delta^2+\vF^2p_x^2}{\hbar^2\omega^2}
\sum\limits_{k=1}^\infty \dfrac{J_k^2\left(2a_0\right)}{k^2} \ll 1 \pnt
$$
\begin{figure*}[t]
	\centering
	\includegraphics[width=1\linewidth]{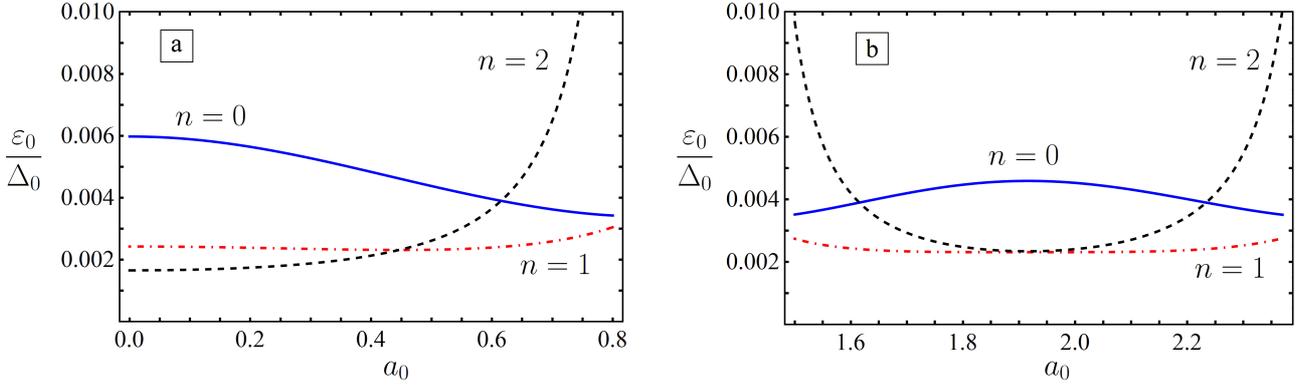}
	\caption{\textbf{Fig. 1.} Binding energy $\varepsilon_0$ vs ac-field amplitude $a_0$. EM radiation is polarized perpendicularly to the SL axis, (a) $g=0.071\Delta_0d$, magnetic field is of 2 T, (b) $g=0.1\Delta_0d$, magnetic field is of 1 T}
	\label{fig1}
\end{figure*}
\\
\\

\textbf{III. Magnetic minibands in ac-driven GSMG}\\

Here we calculate the Floquet spectrum of GSMG in the case of sinusoidal electric field.
The space-periodic modulation of the gap is assumed to be weak. However the presence of the weak gap modulation as well as any other weak space-periodical potential (or SL potential) does not form enough narrow minibands due to the high SL barriers transparency.
As a consequence it can be difficult to achieve the conditions for electron-electron binding. Moreover in the case of weak barriers the system of narrow forbidden bands forms.
So the single-miniband model of SL electron spectrum which is often used for theoretical calculations \cite{50,51,54,55} doesn't work correctly. Another situation we get if we place GSMG in a sufficiently strong magnetic field.
Magnetic field is known to turn the initially continuous spectrum of 2D material into the system of discrete energy levels called as Landau levels.
Eigenvalues and eigenfunctions of Dirac-like Hamiltonian written for electrons in graphene in quantizing magnetic field can be found in \cite{56}, for example.
In the presence of quantizing magnetic field the weak space-periodical potential leads to the expansion of each Landau level into the so-called magnetic miniband \cite{56}.
At that the sufficiently strong magnetic field can provide the situation when minibans are separated by the wide forbidden bands so the single-miniband model can be used.
Note that the forming of the magnetic minibands in the initially gapless graphene placed in the spatial modulated magnetic field had been considered in \cite{57} in the case of absence of ac-field.

Let space-periodic modulation of the gap is created along the axis   and has the profile as follows
\begin{gather}\label{11}
\Delta_\text{s}\left( x\right) =\Delta+\Delta_0\cos\left(\dfrac{2\pi x}{d}\right) \mspace{5mu},\mspace{15mu} \Delta_0\ll\Delta \pnt
\end{gather}

At first the case when EM radiation is polarized perpendicularly to the SL axis $Ox$ is considered.
The inequality presented in (\ref{11}) allows us to use the perturbation theory methods to find the eigenvalues of the Hamiltonian (\ref{8}) which is represented as the sum of two terms
$\hat {\mathscr{H}}_0=\hat {\mathscr{H}}_{\mspace{3mu}\text{f}}+\hat {\mathscr{H}}\mspace{2mu}' \mspace{5mu}$.
Here the term $\hat {\mathscr{H}}_{\mspace{3mu}\text{f}}$ describes the single-layer graphene with gap $\Delta$ in the field with time-independent potential
$\mathbf A\left( \mathbf r\right)$.
The term $\hat {\mathscr{H}}\mspace{2mu}'$ corresponds to the perturbation provided by the weak gap modulation through $Ox$:
\begin{gather}\label{16}
\hat {\mathscr{H}}\mspace{2mu}'=\Delta_0\cos\left(\dfrac{2\pi x}{d}\right)\s{z}\hat\gamma_0 \pnt
\end{gather}

Let GSMG is in the quantizing magnetic field with intensity $\mathbf H$
applied perpendicularly to the graphene plane.
We use the next gauge of the time-independent potential
$\mathbf A\left( \mathbf r\right) =\left(0\mspace{5mu},\mspace{15mu} Hx\mspace{5mu}\right)$.
To find the eigenvalues of unperturbed Hamiltonian we act on the spinor $\chi_0$ twice with the operator
$\hat {\mathscr{H}}_{\mspace{3mu}\text{f}}$.
After using the next relations:
$\s{y}\hat\gamma_0=\hat\gamma_0\s{y}\mspace{5mu}\text{,}\mspace{15mu}\s{x,z}\hat\gamma_0=\hat\gamma_0^+\s{x,z}\com$
we obtain
\begin{multline}\label{1}
-\hbar^2 \vF^2 \hat\gamma_0^+\hat\gamma_0 \mspace{3mu}\dfrac{\text{d}^2\chi_0}{\text{d}x^2}
+\dfrac{\hbar^2\vF^2}{\lambda_H^4}\left( x+\dfrac{\lambda_H^2p_y}{\hbar}\right)^2\chi_0= \\
=\left( \varepsilon_\text{f}^2-\Delta^2\hat\gamma_0^+\hat\gamma_0
-\dfrac{\hbar^2\vF^2}{\lambda_H^2}\s{z}\hat\gamma_0\right) \chi_0 \com
\end{multline}
where $\varepsilon_\text{f}$ is quasienergy of electrons in the absence of the space modulation of the gap,
$\lambda_H=\sqrt{c\hbar/eH}\pnt$
For the sinusoidal ac-field equation (\ref{1}) takes the form
\begin{multline}\label{19}
-\hbar^2 u_\text{\tiny F}^2 \mspace{3mu}\dfrac{\text{d}^2\chi_0}{\text{d}x^2}
+\dfrac{\hbar^2\vF^2}{\lambda_H^4}\left( x+\dfrac{\lambda_H^2p_y}{\hbar}\right)^2\chi_0= \\
=\left( \varepsilon_\text{f}^2-\B{0}{2}{2a_0}\Delta^2
-\dfrac{\hbar^2\vF u_\text{\tiny F}}{\lambda_H^2}\s{z}\right) \chi_0 \pnt
\end{multline}
The eigenfunctions and eigenvalues of (\ref{19}) can be obtained with the use of well known solutions of the harmonic oscillator equation. In the absence of ac-field ($a_0=0$) such solutions are in  \cite{56}, for example.
In our case when the ac-field is present quasienergy is
$\varepsilon_\text{f}\equiv\varepsilon_n=\sqrt{\Delta_\text{ac}^2+\hbar^2\Omega_H^2 n} \mspace{5mu}$,
and eigenfunctions are
$\chi_0=\exp\left(ip_yy/\hbar\right)\left|n\right\rangle \mspace{5mu}$.
Here $\Delta_\text{ac}=\B{0}{}{2a_0}\Delta \mspace{5mu}$,
$\Omega_H=\sqrt{2\left| \B{0}{}{2a_0}\right|}\vF / \lambda_H \mspace{5mu}$,
$$
\left| 0\right\rangle =\dfrac{1}{\sqrt{l_H}}\begin{pmatrix} \mspace{5mu} 0 \mspace{5mu} \\
\mspace{5mu} \Phi_0\left(\dfrac{x-x_H}{l_H} \right)  \mspace{5mu} \end{pmatrix} \com
$$
$$
\left| n\right\rangle =\dfrac{1}{\sqrt{2l_H}}\begin{pmatrix} \mspace{5mu} -i\Phi_{n-1}\left(\dfrac{x-x_H}{l_H} \right) \mspace{5mu} \\ 
\mspace{5mu} \Phi_n\left(\dfrac{x-x_H}{l_H} \right) \mspace{5mu} \end{pmatrix} \com 
$$
$\Phi_n\left(\xi \right)$ is the harmonic oscillator functions,
$x_H=-p_y\lambda_H^2 /\hbar \mspace{5mu}$,
$l_H=\sqrt{\left| \B{0}{}{2a_0}\right| }\lambda_H \mspace{5mu}$.
Using the conditions (\ref{6}) and the formula (\ref{17}) we find that the expressions of electron quasienergy $\varepsilon_n$ and eigenfunctions $\left| n\right\rangle$ derived by averaging over the period of ac-field work correctly if magnetic field, amplitude and frequency of ac electric field obey the next inequality
\begin{gather}\label{15}
\left( \Delta^2+\dfrac{2\hbar^2\vF^2n}{\lambda_H^2}\right) 
\sum\limits_{k=1}^\infty \dfrac{J_k^2\left(2a_0\right)}{\hbar^2\omega^2k^2} \ll 1 \pnt
\end{gather}
For instance, this condition can be easily reached for high frequencies of ac field: $\hbar\omega\gg\varepsilon_n$.

\begin{figure*}[t]
	\centering
	\includegraphics[width=1\linewidth]{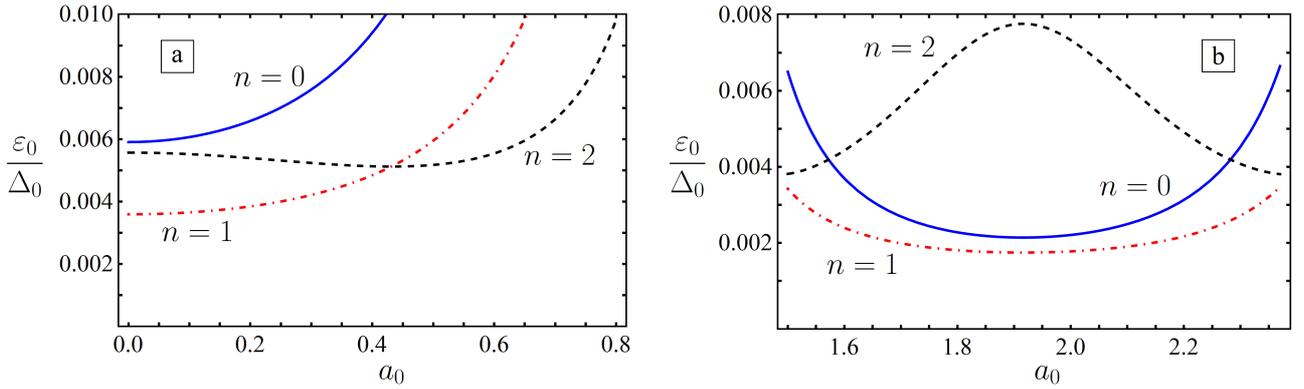}
	\caption{\textbf{Fig. 2.} Binding energy $\varepsilon_0$ vs ac-field amplitude $a_0$. EM radiation is polarized along the SL axis, (a) $g=0.071\Delta_0d$, magnetic field is of 3 T, (b) $g=0.01\Delta_0d$, magnetic field is of 10 T}
	\label{fig2}
\end{figure*}

The presence of weak space modulation of the gap is accounted by the Hamiltonian additional part (\ref{16}).
Such modulation leads to the correction of the quasienergy which is calculated in the first order of the perturbation theory.
These calculations give
\begin{gather}\label{20}
\qe=\varepsilon_n+w_n^\bot\left(a_0\right)\cos{\left(\dfrac{ p_y\Lambda_H}{\hbar}\right)} \com
\end{gather}
where $w_n^\bot\left(a_0\right)=\B{0}{}{2a_0}Q_n\left(\pi l_H / d\right) \Delta_0\mspace{5mu}$,
$\Lambda_H=2\pi\lambda_H^2/d \mspace{5mu}$,
$$
\mspace{-230mu}Q_n\left(\xi \right)=-e^{-\xi^2}\times
$$
$$
\mspace{70mu}\times\begin{cases}
\mspace{15mu} 1 \com \mspace{15mu} n=0 \com \\
\mspace{15mu} \dfrac{L_n\left(2\xi^2\right)-L_{n-1}\left(2\xi^2\right)}{2} \com \mspace{15mu} n\geq 1 \com
\end{cases}
$$
$L_n\left( \xi\right)$ is the Laguerre polynomials.

Thus in the quantizing magnetic field the weak periodic perturbation related with the space modulation of the gap in graphene removes the degeneracy of the electronic states on quasimomentum $p_y$ and each quasienergy level expands into the magnetic miniband.
The width of the $n$-th miniband is equal to $2w_n^\bot$ and it is regulated by the ac-field amplitude changing.
As seen from the above calculations  the electron moves within the miniband as if the SL with the period $\Lambda_H$ was formed through the axis $Oy$.

In the second situation when EM radiation is polarized along the SL axis we derive:
\begin{gather}\label{21}
\qe=\varepsilon_n+w_n^{||}\left(a_0\right)\cos{\left(\dfrac{ p_y\Lambda_H}{\hbar}\right)} \com
\end{gather}
where $w_n^{||}\left(a_0\right)=\B{0}{}{2a_0} Q_n\left(\pi l_H^*/d \right)\Delta_0 \mspace{5mu}$,
$l_H^*=\lambda_H/\sqrt{\left| \B{0}{}{2a_0}\right| } \mspace{5mu}$.
Because of the inequality
$\left| \left\langle n\left|\right.\right.\right.\negthickspace\hat {\mathscr{H}}\mspace{2mu}'\negthickspace
\left.\left.\left.\right| n\right\rangle\right| \ll \varepsilon_n$
is the condition of the applicability of the first-order perturbation theory formulas (\ref{20}) and (\ref{21}) give the correct result if
\begin{gather}\label{24}
w_n^2 \ll \Delta_\text{ac}^2+\hbar^2\Omega_H^2 n \pnt
\end{gather}
Here $w_n$ is chosen to be $w_n^{\bot}$ in the case when ac-field oscillates perpendicularly to the SL axis and it is chosen to be $w_n^{||}$ in the case when ac-field applied along the SL axis.
\\
\\

\textbf{IV. Weakly Bound States of Elementary Excitations in GSMG}\\

In this section we show that in ac-driven GSMG with quasienergy spectrum (19) or (20) the appearance of BS of two electrons is possible at arbitrarily weak repulsion between them. Equation determining the dispersion law of the weak BS $\varepsilon_\text{el}$ can be obtained by calculation the poles of the two-particle Green function \cite{50} and has the form
\begin{multline}\label{22}
\mspace{-10mu}\dfrac{g}{2\pi}\int\limits_{-\pi/\Lambda_H}^{+\pi/\Lambda_H}
\dfrac{\text{d}q}{\bse-\qe\left( q-p/2\right)-\qe\left( q+p/2\right)+i0}= \\
=1 \com
\end{multline}
where $g$ is the effective coupling parameter of quasiparticles with dispersion low $\qe\left( p_y\right) $, $p$ is the momentum of BS.
In the case of repulsion ($g>0$) equation (\ref{22}) has undamped solutions when
$\bse>2\left( \varepsilon_n+w_n\right) $.
After integration in  (\ref{22}) we obtain
\begin{gather}\label{23}
\bse=2\varepsilon_n+\sqrt{\dfrac{g^2}{\Lambda_H^2}+4w_n^2\left(a_0\right)\cos^2\left( \dfrac{p\Lambda_H}{2\hbar}\right) } \pnt
\end{gather}
It is seen from the formula (\ref{23}) that in the case $g\ll \left| w_n\right| \Lambda_H$ electrons can bind if energy $\bse$ is close to the top of conduction band where effective masses are negative.
This fact explains the possibility of electrons binding while $g>0$.
Binding energy is
$\varepsilon_0=g^2 / 4 \left| w_n \left(a_0\right)\right|  \Lambda_H^2$.
The binding energy as the function of the ac-field amplitude $a_0$ in the case when its vector oscillates perpendicularly to the SL axis is shown in \hyperref[fig1]{Fig. 1}.
Dependences shown in \hyperref[fig2]{Fig. 2} correspond to the case when EM radiation is polarized along the SL axis.

When the dependences shown in Figs. 1 and 2 were calculating the next circumstances were taken into account.
The lower limit of the magnetic field was determined by the quantizing conditions and the inequality (\ref{24})
which also ensures the correctness of the perturbation theory method.
The upper limit of the magnetic field was determined by the condition of the weakness of the BS ($g\ll \left| w_n\right| \Lambda_H$).
Besides the last condition arrives at the next requirements.
Firstly, the integer $n$ should not be large ($n=$0, 1, 2) because of the magnetic minibands become too narrow for large $n$.
Secondly, it requires the ac-field amplitude to be far from the roots of the equation $ \B{0}{}{2a_0}=0 $.
\\
\\

\textbf{V. Discussion}\\

Above the Floquet spectrum (quasienergy) of electron in GSMG subjected to the high-frequency EM radiation has been calculated.
The case of the weak space modulation of the gap has been considered.
So to ensure the conditions of the emergence of the electron-electron BS GSMG is placed into the quantizing magnetic field.
The simultaneous presence of the magnetic field and the weak gap modulation leads to the appearance of the magnetic minibands.
The additional ac electric field of high-frequency EM radiation has been shown above to allow the manipulation of the miniband width.
As a consequence it can be used to control the binding energy of electron-electron complex. It is quite seen from the Figs. \ref{fig1} and \ref{fig2}.

In the case of electron-electron coupling BSs correspond to the Bose-Einstein statistics.
Under a critical temperature such electron complexes can form a superfluid gas \cite{58}.
The presence of additional potential in graphene in this case is essential due to the necessity of reaching of the miniband region characterized by the negative effective mass. It provides the electrons binding while $g>0$.
In the absence of additional potential (at the linear dispersion limit) there are no regions with negative effective masses.
So “repulsive” electron-electron interaction can't form bound states in this limit and another mechanisms are necessary to create such BS \cite{49}.
\\
\\

\textbf{Acknowledgements}\\

This work was supported by the RF Ministry of Education and Science as part of State Order no. 2014/411, project code 3154, and within the State Task, code 3.2797.2017/4.6.\\
\end{multicols}
\vspace{5mm}
\begin{center}
	---------------------------------------------------------------------------------------------------------
\end{center}
\vspace{5mm}
\begin{multicols}{2}

\end{multicols}

\begin{thebibliography}{0}
\bibitem{1} F. Molitor, J. Güttinger, C. Stampfer, S. Dröscher, A. Jacobsen, T. Ihn, and K. Ensslin, J. Phys.: Condens. Matter 23, 243201 (2011).
\bibitem{2} H. Fukuyama, Y. Fuseya, M. Ogata, A. Kobayashi, and Y. Suzumura, Physica B 407, 1943 (2012).
\bibitem{3} V.M. Apalkov and T. Chakraborty, Phys. Rev. Lett. 112, 176401 (2014).
\bibitem{4} S.Yu. Davydov, Semiconductors 50, 801 (2016).
\bibitem{5} Z. Sun, T. Hasan, and A.C. Ferrari, Physica E 44, 1082 (2012).
\bibitem{6} G. Konstantatos, M. Badioli, L. Gaudreau, J. Osmond, M. Bernechea, F.P. Garcia de Arquer, F. Gatti, and F.H.L. Koppens, Nat. Nanotechnol. 7, 363 (2012).
\bibitem{7} L. Liao and X. Duan, Mater. Today 15, 328 (2012).
\bibitem{8} D. Bolmatov and C.-Y. Mou, JETP 112, 102 (2011).
\bibitem{9} D.A. Svintsov, V.V. Vyurkov, V.F. Lukichev, A.A. Orlikovsky, A. Burenkov, and R. Oechsner, Semiconductors 47, 279 (2013).
\bibitem{10} A.S. Moskalenko, S.A. Mikhailov, J. Appl. Phys. 115, 203110 (2014).
\bibitem{11} Q. Wilmart, S. Berrada, D. Torrin, V.H. Nguyen, G. Feve, J.-M. Berroir, P. Dollfus, and B. Placais, 2D Materials 1, 011006 (2014).
\bibitem{12} S. Sekwao and J.-P. Leburton, Appl. Phys. Lett. 106, 063109 (2015).
\bibitem{13} V. Gerasik, M.S. Wartak, A.V. Zhukov, and M.B. Belonenko, Mod. Phys. Lett. B 30, 1650185 (2016).
\bibitem{14} M.V. Fistul and K.B. Efetov, Phys. Rev. Lett. 98, 256803 (2007).
\bibitem{15} M. Barbier, P. Vasilopoulos, and F.M. Peeters, Phys. Rev. B 81, 075438 (2010).
\bibitem{16} D.S.L. Abergel and T. Chakraborty, Nanotechnology 22, 015203 (2011).
\bibitem{17} V.M. Apalkov and T. Chakraborty, Phys. Rev. B 86, 035401 (2012).
\bibitem{18} J. Zhu, S.M. Badalyan, and F.M. Peeters, Phys. Rev. Lett. 109, 256602 (2012).
\bibitem{19} Yu.E. Lozovik, and A.A. Sokolik, Nanoscale Res. Lett. 7(1), 134 (2012).
\bibitem{20} H.K. Kelardeh, V. Apalkov, and M.I. Stockman, Phys. Rev. B 90, 085313 (2014).
\bibitem{21} E. Margapoti, P. Strobel, M.M. Asmar, M. Seifert, J. Li, M. Sachsenhauser, Ö. Ceylan, C.-A. Palma, J.V. Barth, J.A. Garrido, A. Cattani-Scholz, S.E. Ulloa, and J.J. Finley, Nano Lett. 14(12), 6823 (2014).
\bibitem{22} S.A. Ktitorov and Chan Siaosin, Tech. Phys. Lett. 36, 433 (2010).
\bibitem{23} N. H. Lindner, G. Refael, and V. Galitski, Nat. Phys. 7, 490 (2011).
\bibitem{24} G. Usaj, P.M. Perez-Piskunow, L.E.F. Foa Torres, and C.A. Balseiro, Phys. Rev. B 90, 115423 (2014).
\bibitem{25} T. Oka and H. Aoki, Phys. Rev. B 79, 081406(R) (2009).
\bibitem{26} Y.H. Wang, H. Steinberg, P. Jarillo-Herrero, and N. Gedik, Science 342, 453 (2013).
\bibitem{27} M. C. Rechtsman, J. M. Zeuner, Y. Plotnik, Y. Lumer, D. Podolsky, F. Dreisow, S. Nolte, M. Segev, and A. Szameit, Nature 496, 196 (2013).
\bibitem{28} S.V. Kryuchkov and E.I. Kukhar’, Physica B 445, 93 (2014).
\bibitem{29} K. Kristinsson, O.V. Kibis, S. Morina, and I.A. Shelykh, Sci. Rep. 6, 20082 (2016).
\bibitem{30} O.V. Kibis, S. Morina, K. Dini, and I.A. Shelykh, Phys. Rev. B 93, 115420 (2016).
\bibitem{31} P. Rodriguez-Lopez, J.J. Betouras, and S.E. Savelev, Phys. Rev. B 89, 155132 (2014).
\bibitem{32} D.A. Lovey, G. Usaj, L.E.F. Foa Torres, and C.A. Balseiro, Phys. Rev. B 93, 245434 (2016).
\bibitem{33} M. Busl, G. Platero, and A.-P. Jauho, Phys. Rev. B 85, 155449 (2012).
\bibitem{34} A. Bostwick, F. Speck, T. Seyller, K. Horn, M. Polini, R. Asgari, A. H. MacDonald, and E. Rotenberg, Science 328, 999 (2010).
\bibitem{35} D. Bolmatov, Physica C 471, 1651 (2011).
\bibitem{36} V.N. Kotov, B. Uchoa, V.M. Pereira, F. Guinea, and A.H. Castro Neto, Rev. Mod. Phys. 84, 1067 (2012).
\bibitem{37} D. Bolmatov and D.V. Zavialov, J. Appl. Phys. 112, 103703 (2012).
\bibitem{38} S. Slizovskiy, Phys. Rev. B 92, 195426 (2015).
\bibitem{39} W.K. Tse and S. Das Sarma, Phys. Rev. Lett. 99, 236802 (2007).
\bibitem{40} D. M. Basko and I. L. Aleiner, Phys. Rev. B 77, 041409(R) (2008).
\bibitem{41} A. Pound, J.P. Carbotte, and E.J. Nicol, Phys. Rev. B 84, 085125 (2011).
\bibitem{42} M.O. Goerbig, J.-N. Fuchs, K. Kechedzhi, and V.I. Fal’ko, Phys. Rev. Lett. 99, 087402 (2007).
\bibitem{43} Y. Kim, J.M. Poumirol, A. Lombardo, N.G. Kalugin, T. Georgiou, Y.J. Kim, K.S. Novoselov, A.C. Ferrari, J. Kono, O. Kashuba, V.I. Fal’ko, and D. Smirnov, Phys. Rev. Lett. 110, 227402 (2013).
\bibitem{44} S. Ihnatsenka and G. Kirczenow, Phys. Rev. B 88, 125430 (2013).
\bibitem{45} J. Jobst, D. Waldmann, I.V. Gornyi, A.D. Mirlin, and H.B. Weber, Phys. Rev. Lett. 108, 106601 (2012).
\bibitem{46} C. Honerkamp, Phys. Rev. Lett. 100, 146404 (2008).
\bibitem{47} S. Pathak, V.B. Shenoy, and G. Baskaran, Phys. Rev. B 81, 085431 (2010).
\bibitem{48} J. González, Phys. Rev. B 78, 205431 (2008).
\bibitem{49} Yu.E. Lozovik and A.A. Sokolik, Phys. Lett. A 374, 2785 (2010).
\bibitem{50} L.P. Pitaevskii, JETP 43, 382 (1976).
\bibitem{51} S.V. Kryuchkov and E.I. Kukhar’, Adv. Condens. Matter Phys. Article ID 979528 (2015).
\bibitem{52} P.V. Ratnikov, JETP Lett. 90, 469 (2009).
\bibitem{53} N.N. Bogoliubov and Y.A. Mitropolsky, Asymptotic Methods in the Theory of Non-Linear Oscillations, Gordon and Breach, New York, (1961).
\bibitem{54} K. Unterrainer, B.J. Keay, M.C. Wanke, S.J. Allen, D. Leonard, G. Medeiros- Ribeiro, U. Bhattacharya, and M.G.W. Rodwell, Phys. Rev. Lett. 76, 2973 (1996).
\bibitem{55} A.V. Shorokhov, M.A. Pyataev, N.N. Khvastunov, T. Hyart, F.V. Kusmartsev, and K.N. Alekseev, JETP Lett. 100, 766 (2015).
\bibitem{56} A. Matulis and F.M. Peeters, Phys. Rev. B 75, 125429 (2007).
\bibitem{57} M. Tahir and K. Sabeeh, Phys. Rev. B 77, 195421 (2008).
\bibitem{58} G.V. Uimin, L.A. Maksimov, and A.F. Barabanov, JETP Lett. 48, 312 (1988).

\end{thebibliography}
\end{document}